\documentstyle[12pt,espcrc1,epsf]{article}  

\newcommand{\AmS}{{\protect\the\textfont2  
  A\kern-.1667em\lower.5ex\hbox{M}\kern-.125emS}}  
  
\title{{}High-$p_\perp$ charged-pion production in Pb-Au Collisions at 158 AGeV/c}  
\author{{} 
G.~Agakichiev$^a$, 
R.~Baur$^b$, 
P.~Braun-Munzinger$^c$, 
A.~Drees$^d$, 
S.~Esumi$^b$, 
U.~Faschingbauer$^{b, e}$, 
Z.~Fraenkel$^f$, 
Ch.~Fuchs$^e$, 
P.~Gl\"assel$^b$, 
C.\thinspace P.~de los Heros$^f$, 
P.~Holl$^g$, 
Ch.~Jung$^b$, 
B.~Lenkeit$^b$, 
F.~Messer$^{d*}$, 
M.~Messer$^b$, 
Y.~Panebrattsev$^a$, 
A.~Pfeiffer$^b$, 
J.~Rak$^e$, 
I.~Ravinovich$^f$, 
S.~Razin$^a$, 
P.~Rehak$^g$, 
M.~Richter$^b$, 
N.~Saveljic$^a$, 
J.~Schukraft$^h$, 
S.~Shimansky$^a$, 
W.~Seipp$^b$, 
E.~Socol$^f$, 
H.\thinspace J.~Specht$^b$, 
J.~Stachel$^b$, 
G.~Tel-Zur$^f$, 
I.~Tserruya$^f$, 
T.~Ullrich$^b$, 
C.~Voigt$^b$, 
C.~Weber$^b$, 
J.\thinspace P.~Wessels$^b$, 
T.~Wienold$^b$, 
J.\thinspace P.~Wurm$^e$, 
V.~Yurevich$^a$ \\ \bigskip  
$^*$~F. Messer: Doctoral Thesis of Federica Ceretto, University of Heidelberg (1998);\\  
$^a$~JINR, Dubna, Russia;   
$^b$~Universit\"at Heidelberg, Germany;   
$^c$~GSI, Darmstadt, Germany;   
$^d$~State University of New York at Stony Brook,USA 
$^e$~Max-Planck-Institut f\"ur Kernphysik, Heidelberg, Germany;   
$^f$~Weizmann Institute of Science, Rehovot, Israel;     
$^g$~Brookhaven National Laboratory, Upton, USA;   
$^h$~CERN, Geneva, Switzerland;  
}   
  
\begin{document}  
\maketitle

\begin{abstract}  
The CERES/NA45 experiment at the CERN SPS measured transverse momentum
spectra of charged-pions in the  range $ 1<p_{\perp}<\rm 4~GeV/c$ near
mid-rapidity ($ 2.1 < y < 2.6 $) in $\rm 158~AGeV/c$ Pb-Au
collisions. The invariant transverse momentum spectra are exponential
over the entire observed range. The average inverse slope is $\rm 245
\pm 5~MeV/c $, it shows a 2.4\% increase with centrality of the
collision over the 35\% most central fraction of the cross section.  The
$\pi^-/\pi^+$  ratio is constant at 1.028$\pm$0.005 over the $
p_{\perp}$ interval measured.  
\end{abstract}   
  
\medskip  
\medskip  
  
Transverse momentum distributions of hadrons which emerge from particle
collisions are closely related to the collision dynamics. In general,
below 1 GeV/c the $p_{\perp}$ spectra exhibit a nearly exponential
slope and reflect the properties of the collision system at break-up,
when the secondary hadrons cease to interact. For example, $p_{\perp}$
spectra of various species measured with lead beam at the SPS have been
used to establish collective flow of hadrons and freeze-out temperatures
in heavy-ion collisions. In pp collisions the spectra
develop a power-law-like tail at much higher $p_{\perp}$ which can be
described quantitatively 
(above about 4 GeV/c) by hard scattering of partons in the
initial phase of the reaction \cite{Owens,Wang,Muller}. In collisions of nuclei
this hard scattering contribution may be strongly modified by partonic
energy loss in a deconfined medium \cite{Gyulassy,Wang}. The
$p_{\perp}$-region  from 1 to 4 GeV/c, often referred to as
semi-hard region, is less easy to interpret. Hard processes begin to 
contribute in this region and compete with low $p_{\perp}$ phenomena.
Additionally, in reactions involving nuclei, the spectra are modified in
the nuclear environment by partons or hadrons scattering more than once
\cite{Lev,Nardi,Becattini}. This was realized by Cronin et
al. \cite{Cronin} at FNAL in the late 70's when in
proton-induced reactions with nuclear targets an enhanced
production of high $p_{\perp}$ particles was found compared to a naive
extrapolation from pp collisions.  Later a similar increase was found
in interactions of $\alpha$-particles at the ISR \cite{faessler} and
also in heavy-ion collisions with oxygen and sulfur beam
at the CERN SPS \cite{helios}. If the hypothesis of a high level of
rescattering, prerequisite for thermalization, is correct, the slope of 
$p_{\perp}$ distributions in the semi-hard region could reflect the
temperature and expansion dynamics of the system.   
 
\bigskip
  
In 1995 the CERES (ChErenkov Ring Electron Spectrometer) experiment has 
taken a data sample of $8\cdot10^6$ Pb-Au collisions at the CERN SPS at
a beam energy of $\rm 158~AGeV/c $. The sample covers the most central 35\%  
of 6230 mb geometrical cross section. In this letter, we
present charged-pion  
transverse momentum spectra derived from $2\cdot10^6$ pions identified in the
range from 1 to 4 $\rm GeV/c$ and near mid-rapidity $ (2.1 < y < 2.6)
$. These data and a preliminary report on hadron spectra has been presented
at QM97 \cite{Ceretto}.  Data on charged-hadron spectra at lower
transverse momenta will be published separately \cite{elsewhere}. 
 
A detailed description of the spectrometer is given in
reference \cite{na45b}. Here we will restrict the discussion of the
experimental setup to aspects essential for the present analysis. 
CERES has been designed to measure electron pairs in ultra-relativistic 
heavy-ion collisions. Two RICH detectors, with full azimuthal coverage,
located before and after a superconducting double solenoid constitute
the heart of the experiment. The magnet system deflects each track in
azimuthal direction, but maintains the polar. Precise tracking is
provided by two silicon drift detectors (SDD) positioned on average 10 cm and 11.5 cm
downstream of the target in front of the first RICH and a pad chamber
(PC) downstream of the second RICH. The two SDDs also determine the
position of the event vertex and measure the charged-particle density
$\rm dN_{ch}/d\eta$. By choosing CH$_4$ as the radiator gas for the 
RICH detectors the Cherenkov threshold   
is high enough ($\rm \gamma_{th} \approx 32$) to suppress signals from
the bulk of hadrons. However, 
charged pions above 5 GeV/c momentum radiate enough  Cherenkov light to
be observed in the RICH detectors. The clean environment necessary to
detect electrons among hundreds of hadrons is  ideally suited for a
high-statistics study of those pions which exceed the Cherenkov
threshold. Pion identification and momentum measurement are provided 
via the ring radius measured in the RICH detectors. The azimuthal
deflection in the magnetic field gives charge information and redundant
momentum determination according to $\rm \Delta\phi=144~mrad/{\it p}(GeV/c)$. 

\bigskip

The analysis proceeds in several steps. In the first step pion
candidates are reconstructed. Each charged particle track originating
from the vertex and traversing the two SDDs is extrapolated downstream
to the Pad Chamber. If a Pad Chamber signal is found in a  window of
$\rm 6~mrad$ in polar and $\rm 50~mrad$ in azimuthal direction
(corresponding  to a lower momentum cut  of $\rm 2~GeV/c$) ring images
are searched for in both RICH detectors at the corresponding
locations. Several constraints are imposed to identify a pion track: (i)
the two ring images match in radius within 10\%, (ii) the radii correspond to
the azimuthal deflection in the field within two sigma of the resolution,
and (iii) the polar angle coincides within two sigma of the measured resolution
in all detectors. Tight cuts on using these
constraints remove most false  tracks from the sample.   
 
\begin{figure}[htb]  
 \includegraphics{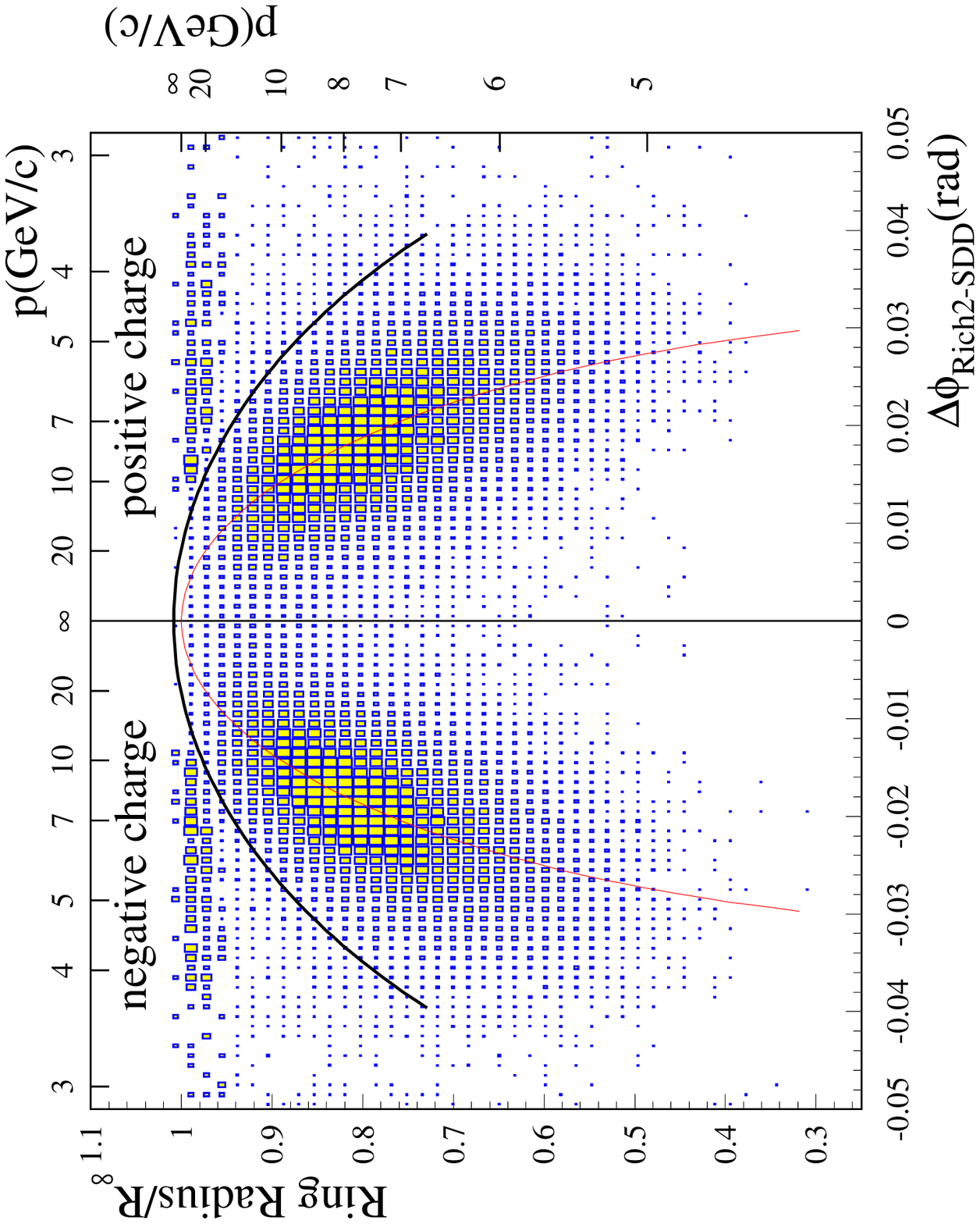}  
  \vspace{10.cm}  
    \caption{{} Correlation between Cherenkov ring radius (in units of
        the asymptotic ring radius  $R_\infty$) and azimuthal
        deflection in the magnetic field for  reconstructed positive and
        negative pions. The top and right side of the figure indicate
        the corresponding momentum scales derived from both quatities. 
        The dotted line shows the correlation expected for
        pions. The solid line represents the cut used to suppress
        electron tracks which survive the two sigma cut on the match
        between radius and phi deflection.
        In order to illustrate the cut and to visualize
        this background,  which is at a very low level, all entries
        above  $R/R_\infty$ of 0.95 have been multiplied by a factor
        20.\label{rdphi}}   
\end{figure}

Fig.~\ref{rdphi} depicts the correlation of ring radius and azimuthal
deflection for all the reconstructed pion candidates after applying the
cuts (i) and (iii). Two well-defined regions, corresponding to pions
of negative and positive charge, cluster around the expected
correlation (dotted line) and clearly stand out above a low level of
remaining background tracks.  
 
Nearly all remaining background tracks in the candidate
sample can be associated with electron tracks. Two classes of background
tracks must be distinguished: (i) physical background of high-momentum
electrons and (ii) unphysical background originating from uncorrelated
electron rings accidentally matching the track in both RICH detectors.
High momentum electrons are easily identified by the mismatching
of ring radius, which always is the asymptotic one, and 
deflection in the magnetic field. The cut used to remove those tracks is
depicted in the figure (solid line). Note that this background was
scaled up by a factor of 20 to make it visible in the figure. The electron
background in the remaining sample was determined using a Monte-Carlo
simulation and was found to be negligible at all momenta. The unphysical
combinatorial background was measured using an event-mixing technique
and subtracted from the data. The low level of this background  may be
judged from the amount of tracks in Fig.~\ref{rdphi} with uncorrelated
radius and deflection. The signal-to-background ratio is 100 at a $p_\perp$
of 1~GeV/c, and decreases to 10 above 4~GeV/c. 
At the highest $p_\perp$ the background subtraction introduces a
systematic error of less than 5\%. Kaons are only reconstructed if their
momentum is above 20 GeV/c, thus kaon background is suppressed by four
order of magnitude at all $p_\perp$. 

The momentum can be calculated from the azimuthal deflection and the
ring radius, both  scales are shown in Fig.~\ref{rdphi}. In our analysis
we use the momentum based  on the ring radius which has a resolution of
${\Delta p}/{p} \sim 0.0008\cdot p^2 \rm(GeV/c)^2$, significantly better
in the range from  4.5 to 30 $\rm GeV/c$ than the resolution $ {\Delta
p}/{p} \sim 0.035 \cdot p\rm(GeV/c) $ obtained from the deflection in
the magnetic field. The comparison of both momentum measurements fixes
the absolute momentum scale to better than 0.5\%. 

\begin{figure}[htb]  
 \includegraphics{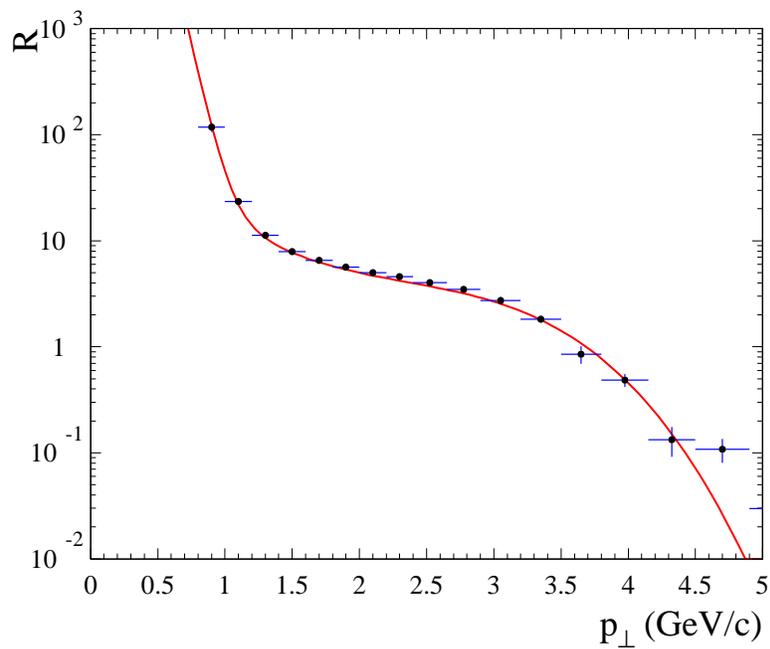}  
  \vspace{10.cm}  
    \caption{{} Correction function used to correct the pion $p_{\perp}$
        distribution obtained for the full data set. The function is
        determined iteratively from a Monte Carlo simulation as ratio 
        of input distribution to the one reconstructed through the full
        analysis chain. The function corrects for limited reconstruction
        efficiency, all analysis cuts, acceptance and resolution of the
        spectrometer. \label{cf}}  
\end{figure}

In the last step of
the analysis the measured transverse momentum spectrum is corrected for
the spectrometer characteristics, the limitations of the reconstruction
algorithm, as well as all cuts on the data. All corrections are
calculated simultaneously in a  Monte-Carlo simulation.  Pions,
generated with realistic kinematical distributions, were  traced through
a detailed GEANT \cite{geant} implementation of the CERES spectrometer.  The pion
tracks, subjected to the simulated response of the detector, are
embedded in real events and passed through the  full analysis chain. The
ratio $ R$ of Monte-Carlo input to output $ p_{\perp}$ distributions  is
shown in Fig.~\ref{cf}. The fit to the ratio (as shown in Fig.~\ref{cf})
is used to correct the
data.   
 
The correction function can be split into three regions. Below
$p_{\perp}$ of 1.3~GeV/c the correction increases rapidly as pions fall
below the Cherenkov threshold. Expressed in $p_{\perp}$, the threshold
depends on the polar angle and hence the cut-off in Fig.~\ref{cf} is
gradual. The plateau region  up to $\rm 3~GeV/c$  reflects the
reconstruction efficiency of about 20\% including all losses due to
analysis  cuts. Above the Cherenkov threshold the number of photons and
the ring radius steadily increase towards their asymptotic values. Both
yield an increase of the reconstruction efficiency and the ring center
resolution with increasing momentum, which results in a gradual reduction of
the correction towards higher momenta. Above $\rm 3~GeV/c$ the
correction factor decreases more rapidly due to  the deterioration of
the momentum resolution which artificially increases the apparent yield
at high momenta.  The systematic uncertainty in the region from 1.5 to 3
$\rm GeV/c$ is less  than 10\%, below 1.5 $\rm GeV/c $  the error
increases slightly up to 15\% at 1.2 GeV/c. The correction for momentum
smearing generates uncertainties up to  40\% at 4 $\rm GeV/c $. The
absolute normalization introduces an additional overall systematic
error of 20\% independent of $p_{\perp}$.  
 
Fig.~\ref{cf} shows the correction function for the full data set,
corresponding to the upper 35\% of the geometrical cross section. To study
the centrality dependence the data were split into 7 exclusive multiplicity
bins, covering the range from $dN_{ch}/d\eta$ of 100 to about 400. For
each bin the correction was calculated separately to 
take into account the reduction of the reconstruction efficiency and 
the deterioration  of the momentum resolution with  event multiplicity.
The absolute value of the correction function increases by a factor of 2
from the lowest to the highest multiplicity events. No change of the
$p_{\perp}$ dependence of the correction was observed.
 
\begin{figure}[htb]  
\includegraphics{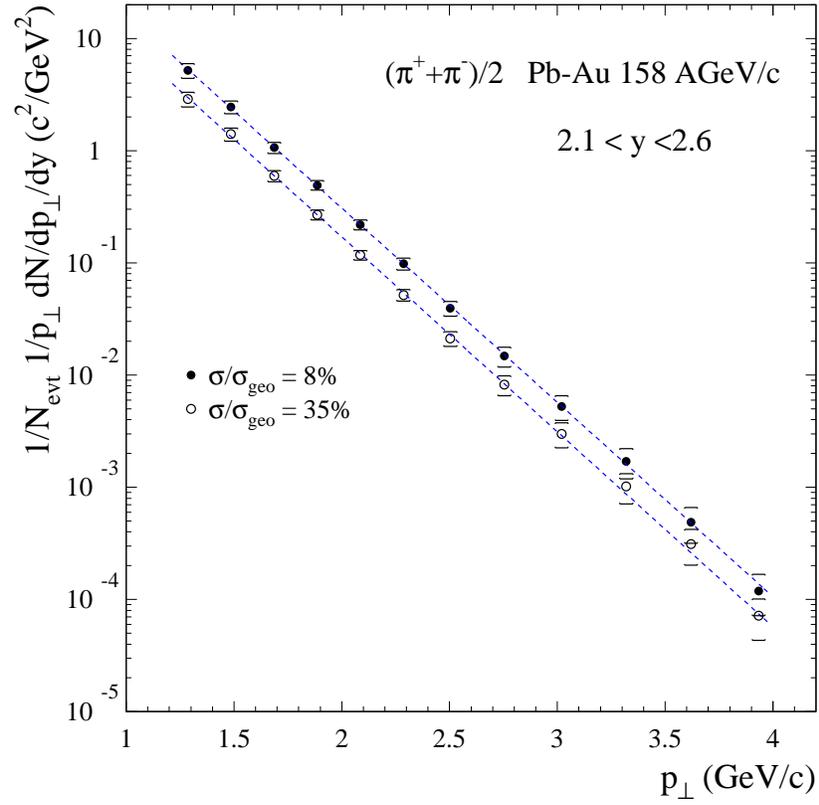}   
  \vspace{10.cm}  
    \caption{{} Invariant transverse momentum distribution of positively and
        negatively charged pions. Shown are the distributions for the full
        data sample and for the most central fraction of the data,
        corresponding to 35\% and 8\% of the geometrical cross section,
        respectively. The lines are exponential fits to the data.
        \label{spectra}}  
\end{figure}

The final  $p_{\perp}$ spectrum is shown in Fig.~\ref{spectra}. Since
$\pi^+$  and $\pi^-$ spectra are very similar (see the discussion below)
we have averaged positively  and negatively charged pions in order to
cover the  largest possible $p_{\perp}$ range. The open symbols
represent the full data sample with an average charged-particle density
of 220 corresponding to the upper 35\% of the geometrical cross section.
The full symbols give the result for a more central event selection,
with  $\rm dN_{ch}/d\eta=310$ or about 8\% of the geometrical cross
section. Our result is in good agreement with the recently measured
$\pi^\circ$ spectrum in  Pb-Pb collisions at the same energy and similar
rapidity \cite{wa98}.     
 
Except for the increase in the yield proportional to the charged
particle density both spectra shown in Fig.~\ref{spectra} are 
indistinguishable. The spectra are exponential over
the full range observed. Fitting an exponential function 
$Ae^{(-p_{\perp}c/T)}$  to the full data sample in the range 1.5 to 3.5
$\rm GeV$ gives an inverse slope parameter T  $\rm = 245 \pm 5~
MeV/c$. The fit of the $m_\perp-m_\circ$ distribution gives within 1 MeV
the same inverse slope parameter. 
The inverse slope varies by less than 10 MeV locally over 
the full $p_{\perp}$ range. Note that 
the slope parameter is significantly larger than  the values of about 
$\rm  180~MeV$ which have been observed at lower $p_{\perp}$
\cite{Ceretto}.   
 
\begin{figure}[htb]  
 \includegraphics{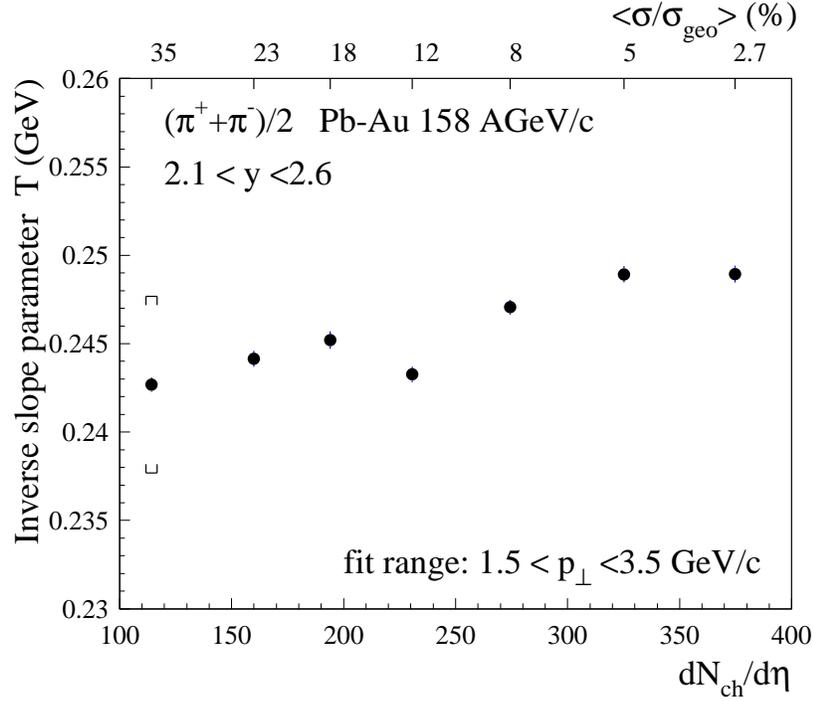}  
  \vspace{10.cm}  
    \caption{{} Centrality dependence of the inverse slope parameter
        T. The slope parameter is extracted from fitting an exponential
        function to the invariant $p_\perp$ distribution obtained from
        exclusive event samples. The abscissa shows the average charged
        particle density of each event sample. The scale at the top
        gives the fraction of the geometrical cross section integrated 
        above the coresponding charged particle density. Error bars 
        represent statistical erros, the bracket indicates the
        systematic error.  
        \label{slope}}  
\end{figure}

We have split the data sample into 7 exclusive multiplicity bins. The
inverse slope parameter T extracted from each of these bins is plotted
in Fig.~\ref{slope} versus the average $\rm dN_{ch}/d\eta$ of the bin.
Over the measured centrality range the inverse slope increases by 7
MeV,  which is about 2.4\% of its absolute value.
The statistical errors are smaller than 1 MeV. The systematic error
is 5 MeV (indicated by the brackets in Fig.~\ref{slope}) nearly
independent of centrality, thus it is an error 
on the absolute value but not on the trend of the slope with centrality. 

\begin{figure}[htb]  
 \includegraphics{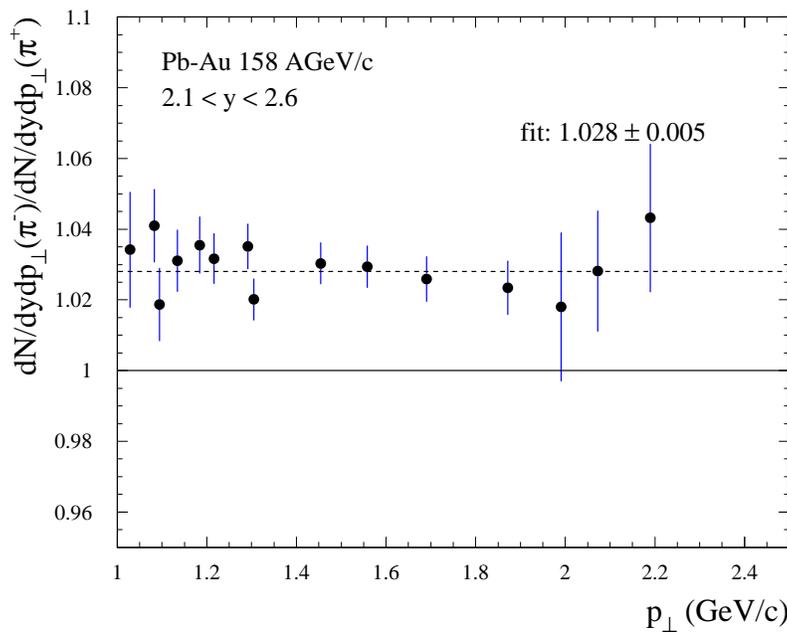}  
  \vspace{10.cm}  
    \caption{{} Transverse momentum dependence of the 
        $\pi^/\pi^+$ ratio. \label{ratio}}   
\end{figure}

From Fig.~\ref{rdphi} it is evident that $\pi^+$ and $\pi^-$ can be
measured separately. We have determined the $\pi^-/\pi^+$ ratio for 
15 exclusive ring radius bands. The average momentum corresponding to
each radius bin is determined using the CERES Monte-Carlo simulation. 
In particular for large $p_\perp$, this momentum is smaller than
calculated from the average ring radius because low momentum pions 
shift to apparently larger momentum due to the limited resolution. 
As discussed previously, the inclusive $p_\perp$ spectra were corrected
for the effect of the momentum resolution. While the systematic errors
introduced by this correction are acceptable for the inclusive spectra
up to 4 GeV/c, they exclude a meaningful comparison of  $\pi^-$ to
$\pi^+$  above 2.2 GeV/c. The transverse momentum dependence of the
$\pi^-/\pi^+$ ratio is shown in Fig.\ref{ratio}. The $\pi^-/\pi^+$ ratio
is constant within errors over the entire observed range at a value of 
1.028 $\pm$ 0.005, assuming that the systematic errors cancel in the ratio.

\bigskip

The $p_\perp$ spectra of charged pions are nearly exponential over more than
four orders of magnitude from 1 to 4 GeV/c in $p_\perp$. This suggests a
statistical interpretation of the data. Of course, the inverse slope of
about 245 MeV, well above the Hagedorn temperature of 160 MeV \cite{Hagedorn},
can not be 
interpreted as temperature of a dense system of hadronic
resonances. This could point towards early thermalization in a partonic
phase. On the other hand, collective transverse expansion, well
established by observing a linear increase of the inverse slope with
particle mass at lower transverse momenta, might sufficiently increase
the inverse slope even at the large $p_\perp$ observed in this
experiment. Data on $\pi^\circ$ production at similar $p_\perp$ have been
interpreted accordingly \cite{wa98}. 

Models with initial state scattering on the hadron or parton level
\cite{Lev,Nardi} can explain the 
momentum spectra for central collisions
\cite{Ceretto,Wang}. However, such ``random walk'' models lead to a much
larger centrality dependence of the slope and can be excluded \cite{pbm}. 
Wang has shown \cite{Wang} that perturbative QCD
calculation modeling the Cronin effect by $p_\perp$ broadening can
explain $\pi^\circ$ data down to 1 GeV/c, and one wonders why they do
not seem to reflect any parton energy loss in the hot and dense medium. 
However, the results depend sensitively on the model of the Cronin
effect \cite{gyu}. One might 
speculate that the number of consecutive parton scattering processes will 
increase as the impact parameter decreases. 

The $\pi^-/\pi^+$ ratio, which reflects the isospin asymmerty in the
final state, is different for statistical and perturbative QCD based 
models. Perturbative QCD predicts a $\pi^-/\pi^+$ ratio increasing with $p_\perp$ and 
saturating at 1.14, which corresponds to the initial isospin unbalance 
of the valece quarks. The constant ratio at a much lower
level observed in the experiment indicates that below 2.2 GeV/c hard
scattering does not dominate. A statistical model which evenly
distributes the initial isospin unbalance over all particles in the
final state gives a ratio of about 1.06. A more sophisticated analysis,
including the effects of hadron decays in the final state
\cite{Stachel}, results in 
ratio of 1.05. It increases  below 400 MeV/c $p_\perp$ but remains fairly
constant at larger $p_\perp$.  

It remains ambiguous whether a statistical or a perturbative QCD 
interpretation of the data is more appropriate. Additional information
might arise from the observation of angular correlations between high
momentum pions. 

\medskip
\medskip
We are grateful for the financial support by the MINERVA Foundation and the
Benoziyo High Energy Research Center.

\vspace{3.5cm}

\end{document}